\documentclass[prb,twocolumn,groupedaddress]{revtex4-1}
\usepackage{amssymb}

\usepackage{graphicx}
\usepackage{epsfig}
\usepackage{latexsym}
\usepackage{amsmath}
\usepackage{bm}
\usepackage{color}
\usepackage{bbold}

\begin{document}

\title{The Hall Effect in Superconducting Films}

\begin{abstract}
Near the superconducting phase transition, fluctuations significantly modify the electronic 
transport properties. Here we study the  fluctuation corrections to the Hall conductivity in disordered films, 
extending  previous derivations to  a broader range of temperatures and
magnetic fields, including the vicinity of the magnetic field induced quantum critical point. In the process, we found a new contribution to the Hall conductivity that
was not considered before. Recently, our theory has been used to fit measurements of the Hall resistance in amorphous TaN films. 
\end{abstract}

\author{Karen Michaeli}
\affiliation{Department of Physics, Massachusetts Institute of Technology,
77 Massachusetts Avenue, Cambridge, MA 02139}
\email{karenmic@mit.edu}
\author{Konstantin S. Tikhonov}
\affiliation{ Department of Physics, Texas A\&M University, College Station, TX $77843-4242$, USA}
\author{Alexander M. Finkel'stein}
\affiliation{Department of Condensed Matter Physics, The Weizmann Institute of Science,
Rehovot 76100, Israel  and \\ Department of Physics, Texas A\&M University, College Station, TX $77843-4242$, USA}

\maketitle

Measurements of the Hall effect in the classically weak magnetic fields
provide useful information about the density of the current carriers as well
as the sign of their charge. According to the Drude formulas, the ratio
between the Hall ($\sigma_{xy}$) and longitudinal ($\sigma_{xx}$)
conductivities is $\omega_c\tau$, where $\omega_c=|eH/m^{*}c|$ is the
cyclotron frequency of the quasiparticles (electrons or holes) and $\tau$ is
the elastic scattering time. The appearance of the cyclotron frequency in
the expression for $\sigma_{xy}$ manifests the fact that for the Hall effect
to be finite particle-hole asymmetry is required. As is well known, within
the Drude model the Hall coefficient is independent of $\tau$ and $\omega_c$%
, and is only function of the charge carriers density $n$; $R_{H}\equiv
\rho_{xy}/H=1/nec$. Weak localization corrections arising due to the
interference effects although modifying both $\sigma_{xy}$ and $\sigma_{xx}$
leave $R_H$ unchanged. In contrast, electron-electron interactions affect
the transverse and longitudinal components of the conductivity tensor in a
way violating the delicate balance between them and, therefore, $R_H$ is no
longer universal. In particular, a significant change in the Hall
coefficient occurs near the superconducting transition as a result of the
fluctuations induced by electron-electron interaction in the Cooper channel.
As we show here, the corrections to the Hall conductivity due to
superconducting fluctuations diverge stronger than the longitudinal ones.
Furthermore, the particle-hole asymmetry factor $\omega_c\tau$ is multiplied
by $\varsigma\mu$ that makes it parametrically larger. The parameter $%
\varsigma$ is proportional to the derivative of the density of states with
respect to the energy at the chemical potential $\mu$. The only other
transport property that is sensitive to this quantity is the thermoelectric
coefficient.~\cite{Mott}

Close to the superconducting phase transition, yet in the normal metallic
phase, the fluctuations of the superconducting order parameter form a new
branch of collective excitations. Since these excitations are charged, they
create a new channel for the electric current. As a result, the electric
conductivity is determined not only by the single-particle excitations
(quasiparticles), but also by the current carried by the fluctuations. The
direct contribution of the superconducting fluctuations to the longitudinal
electric conductivity is described by the Aslamazov-Larkin term.~\cite{Aslamazov1968} In the vicinity of
the transition, this contribution can be interpreted as the Drude
conductivity of the fluctuating Cooper pairs. Besides, the fluctuations affect strongly the quasiparticles, and by
that influence the conductivity. The scattering of the current-carrying
quasiparticles by the superconducting fluctuations are described by the
Maki-Thompson term.~\cite{Maki1968,Thompson1970} Another effect can be
attributed to the modification of the quasiparticles density of states by
the long living superconducting fluctuations.~\cite{Varlamov}

Similar to the Hall conductivity of free electrons, the corrections to $%
\sigma_{xy}$ generated by the superconducting fluctuations vanish in the
absence of particle-hole asymmetry. To demonstrate the dependence of the
conductivity on the particle-hole asymmetry, we shall use the
Aslamazov-Larkin corrections as an example. Close to $T_c$, the
superconducting fluctuations can be described using the time dependent
Ginzburg-Landau (TDGL)~\cite{schmid1966,abrahamsts1966,gorkoveliash1968} equation:
\begin{align}  \label{eq:TDGL1}
-\frac{a}{T_{c}}&\left( \frac{\partial }{\partial {t}}+2ie\varphi \right)
\Delta (\mathbf{r},t)\\\nonumber&=\left[ \frac{T-T_{c}}{T_{c}}+\frac{\pi D}{8T_{c}}
\left( -i\boldsymbol{\nabla}-2e\mathbf{A}\right) ^{2}\right] \Delta (\mathbf{%
r},t).
\end{align}
Here $\Delta(\mathbf{r},t)$ is a complex field describing the order
parameter fluctuating in time and space (a detailed discussion of the
TDGL-theory can be found in Ref.~\onlinecite{Varlamov}). The coefficient $a$
is known from microscopic calculations to be equal to $\pi/8$, and $e=-|e|$
is the electron charge. The first term on the right hand side corresponds to
the finite energy needed to create a fluctuation of the superconducting
order parameter above the transition temperature. We can look at the
semi-phenomenological equation presented in Eq.~\ref{eq:TDGL1} as describing
$2e$-charged particles with a life-time $\tau_{\Delta}\sim(T-T_c)^{-1}$. The
conductivity associated with these particles is simply their Drude
conductivity, $\sigma_{xx}=(2e)^2n_{\Delta}\tau_{\Delta}/m_{GL}\sim{e^2T}%
/(T-T_c)$. A comparison with the microscopic calculations shows that the
Aslamazov-Larkin contribution to the longitudinal conductivity coincides
with the one obtained using the semi-phenomenological equation. However, no
correction to the Hall conductivity can be generated as far as the dynamics
of the superconducting fluctuations remains within the form given by Eq.~\ref%
{eq:TDGL1}.

The TDGL-equation can be derived directly from the microscopic theory by
integrating out the single-particle degrees of freedom. Then, under the
assumption that the quasiparticles have a constant density of states, one
arrives to Eq.~\ref{eq:TDGL1}. Since no particle-hole asymmetry has been
introduced, the excitations associated with the superconducting
fluctuations, as described by Eq.~\ref{eq:TDGL1}, are invariant under
particle-hole transformation. Therefore, it should not be surprising that
the contribution of the superconducting fluctuations to the Hall
conductivity vanishes in the framework of this equation.  It has been first
pointed out by Fukuyama \textit{et al.}~\cite{Fukuyama1971} that the
Aslamazov-Larkin correction vanishes unless the derivative of the density of states with respect to the energy is taken into account. In other words, this contribution to the Hall conductivity depends on the particle-hole asymmetry. This important observation was the basis for subsequent studies of the Hall effect in the framework of TDGL theory both for conventional and high-$T_c$ superconductors  as well as in the flux-flow
regimes.~\cite{Dorsey,UllahDorsey,Kopnin,feigelman1995,Angilella}.

Aronov \textit{et al.}~\cite{Larkin1995,Aronov1992} incorporated the particle-hole asymmetry into the TDGL equation by adding a new term:
\begin{align}  \label{eq:TDGL2}
-\left( \frac{\partial }{\partial {t}}+2ie\varphi \right)& \left( \frac{a}{%
T_{c}}+i\varsigma \right) \Delta (\mathbf{r},t)\\\nonumber&\hspace{-10mm}=\left[ \frac{T-T_{c}}{T_{c}}+%
\frac{\pi D}{8T_{c}} \left( -i\boldsymbol{\nabla}-2e\mathbf{A}\right) ^{2}%
\right] \Delta (\mathbf{r},t).
\end{align}
This equation was used to derive the Aslamazov-Larkin correction to the Hall conductivity. The authors of Ref.~\onlinecite{Larkin1995} claimed that the new parameter,
can be related to the derivative of the critical temperature
with respect to the chemical potential, $\varsigma=-0.5d\ln T_{c}/d\mu \sim
-\lambda^{-1}\nu^{\prime}(\mu)/\nu(\mu)$. Here $\lambda $ is the dimensional
coupling constant determining $T_{c}=\omega_{D}\exp (-1/\lambda )$, and $%
\nu(\mu)$ is the density of states at the Fermi energy while $%
\nu^{\prime}(\mu)$ is its derivative with respect to the energy.  Hence, the corrections to the
Hall conductivity, being proportional to $\varsigma$, can provide
information on the dependence of the density of states on the energy.
Microscopic calculation presented in Appendix~\ref{sec:asymmetry} confirms that for three dimensional electrons $\varsigma$ is proportional to $1/(\lambda \varepsilon _{F})$. [Throughout the entire paper we consider a not too thin film in which the electrons are three dimensional while the superconducting fluctuations are two dimensional.]
The analysis of Eq.~\ref{eq:TDGL2} reveals that in the diffusive regime the
cyclotron frequency corresponding to the charged field $\Delta$ is equal to $\Omega_c=|4eHD/c|$, where $\Omega_c\propto(\varepsilon_F\tau)\omega_c\gg%
\omega_c$. In $\Omega_c$, the effective charge is equal to $2e$ and the
diffusion coefficient $D$ replaces $1/2m$, because in the fluctuation
propagators the kinetic energy $p^2/2m$ is substituted by $Dq^2$.
Consequently, the Drude-like contribution of the superconducting
fluctuations to the Hall conductivity is proportional to $\varsigma\Omega_c$.

In this paper we extend previous theoretical analysis~\cite{Fukuyama1971,Larkin1995,Aronov1992} of the the corrections to the Hall conductivity
for various temperatures and magnetic fields. Although the diagonal component of the magnetoresistance has been studied for the entire phase diagram including the vicinity of the Quantum Critical Point, induced by magnetic field~\cite{Galitski2001}, up to now there was no similar systematic analysis of the Hall resistance. The results for the leading corrections to the Hall conductivity generated by the superconducting fluctuations are summarized in Fig.~\ref{fig:PhaseDiagramHall}. This work has been inspired by recent
measurements of the Hall conductivity in  disordered Tantalum Nitride films.~\cite{Kapitulnik} Some of the results presented here have been used in Ref.~\onlinecite{Kapitulnik} for the analysis of the Hall conductivity measurements.

As we explained above, the particle-hole asymmetry enters the Hall conductivity either via the quasiparticle mass (or equivalently, the cyclotron frequency $\omega_c$) or the derivative of the density of states. While the former appears when the Lorentz force acts on the quasiparticles in order to turn the current from the longitudinal to the transverse direction, the latter appears when the Lorentz force acts on the superconducting fluctuations. Thus, in general, there are two distinct types of corrections to the Hall conductivity, one proportional to $\omega_c\tau$ and the other to $\varsigma\Omega_c\sim\omega_c\tau/\lambda$. Since the coupling constant for the superconducting interaction is usually much smaller than unity, one may expect only the second kind of contributions to be important. However, the two contributions also differ in their dependence on the distance from the superconducting transition, $\ln{T/T_c(H)}$ or $\ln{H/H_{c2}(T)}$. Moreover, we have found a new term which, although is not enhanced by the inverse coupling constant $1/\lambda$, contributes to the transverse conductivity in a broad range of temperatures and magnetic fields. In particular, this contribution, unique to the Hall conductivity, gives the most dominant fluctuation correction to $\sigma_{xy}$ far from the transition at $T\gg{T_c}$.

The rest of the paper is organized as follows: The derivation of the Hall conductivity using the quantum kinetic equation is discussed in Section~\ref{sec:QKE} and Appendix~\ref{Derivation}. The results of the calculation for the different regions of the $T$-$H$ phase diagram are given in Section.~\ref{sec:phasediagram}.

\section{Derivation of the Hall conductivity}

\label{sec:QKE}

\begin{figure}[tp]
\begin{flushright}\begin{minipage}{0.5\textwidth}  \centering
      \includegraphics[width=1\textwidth]{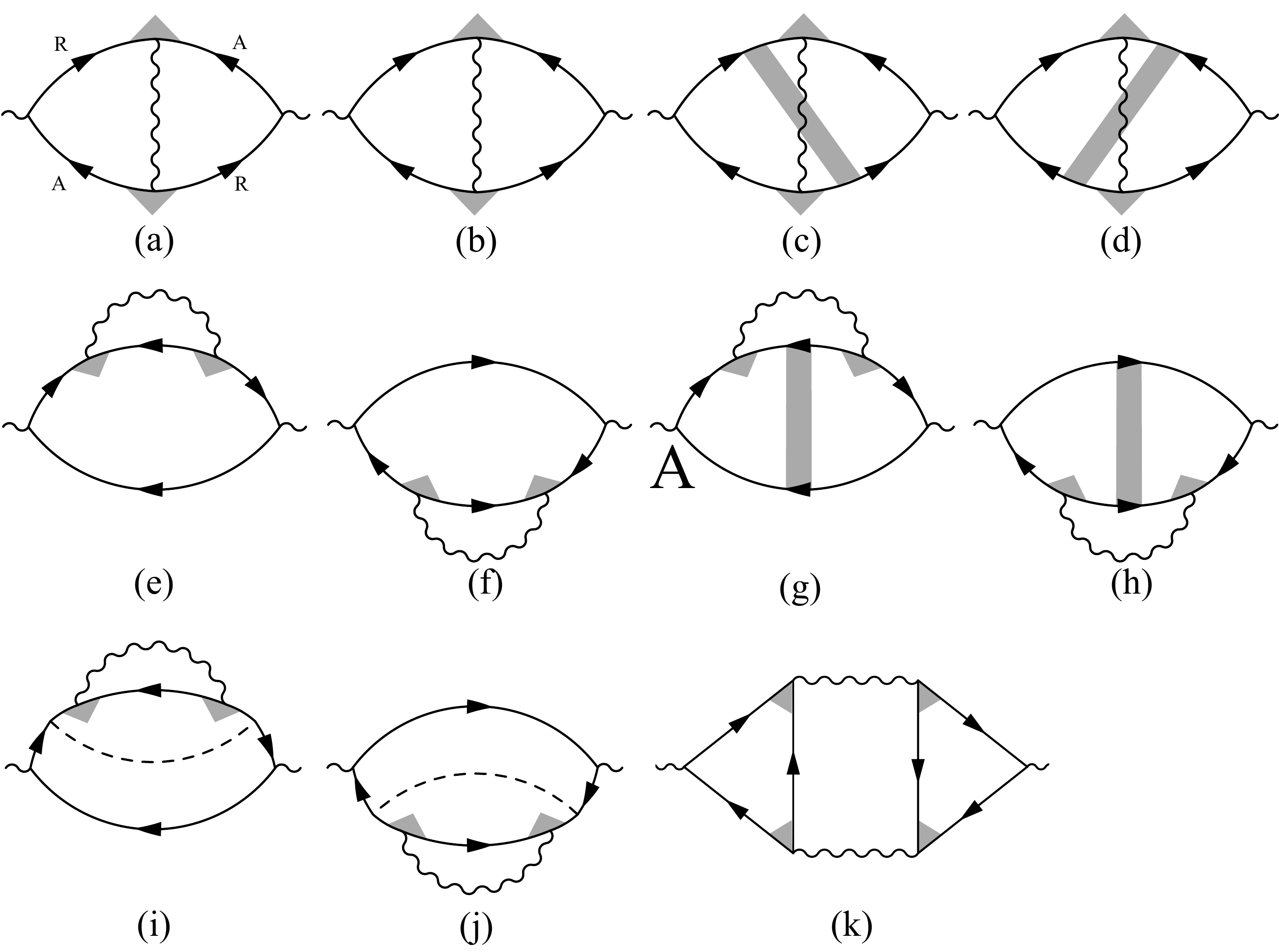} \hspace{0.05in}
               \caption[0.4\textwidth]{\small The eleven diagrams contributing to the superconducting fluctuations corrections to the longitudinal conductivity $\delta\sigma_{xx}$. a. The anomalous Maki-Thompson corrections. The analytical structure of the different Green's functions are indicated by R (retarded) and A (advanced). b-d. The regular Maki-Thompson corrections. e-j. The density of state corrections. k. The Aslamazov-Larkin term.}
               \label{fig:diagrams}
\end{minipage}\end{flushright}
\end{figure}

\begin{figure}[tp]
\begin{flushright}\begin{minipage}{0.5\textwidth}  \centering
      \includegraphics[width=1\textwidth]{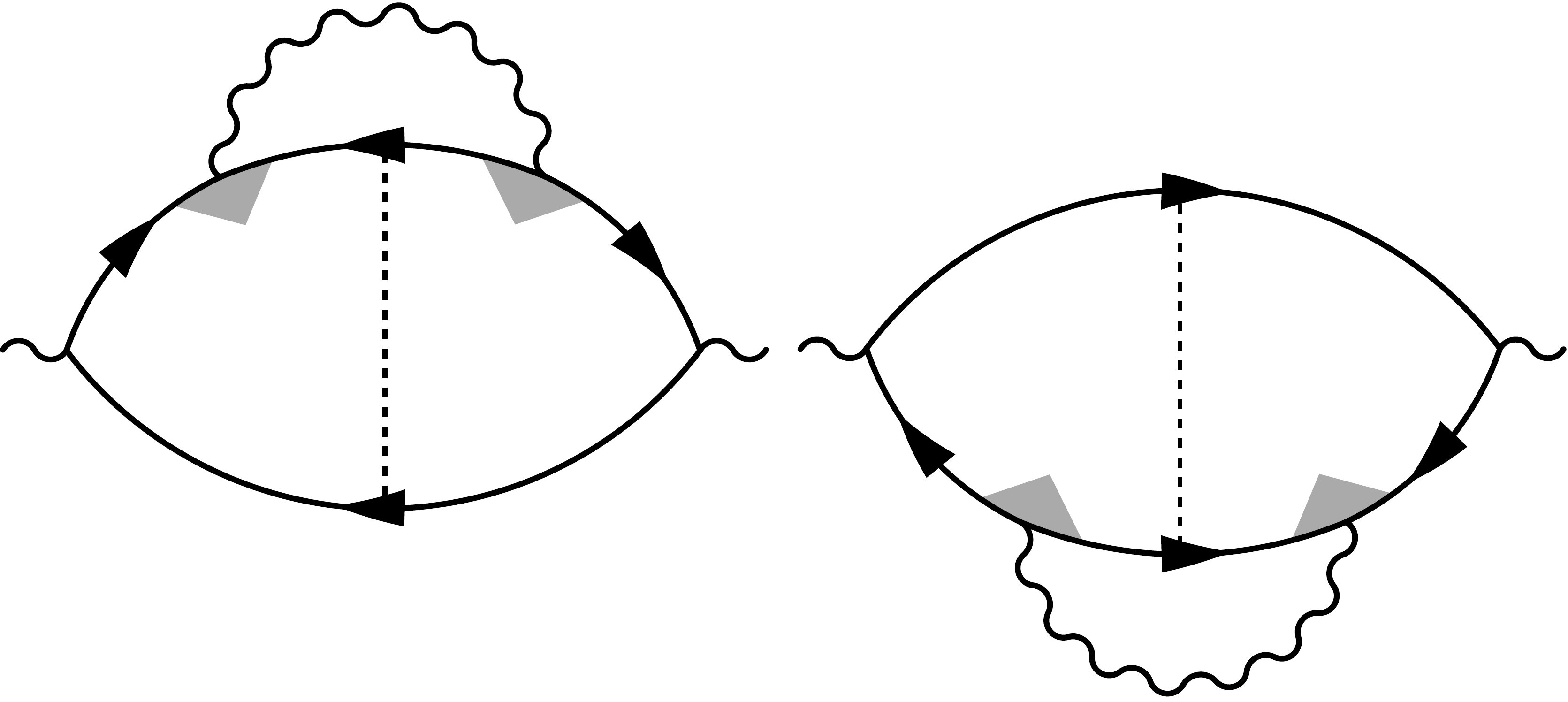} \hspace{0.05in}
               \caption[0.4\textwidth]{\small The new contribution to the Hall conductivity.}
               \label{fig:new1}
\end{minipage}\end{flushright}
\end{figure}

For the derivation of the Hall conductivity we apply the quantum kinetic
technique,~\cite{NernstLetter,Nernst,QKE} but the same result can be
obtained using the Kubo formula. The details of the derivation are described
in Appendix~\ref{Derivation}. For the purpose of illustration, we use diagrammatic
representation for the different contributions to the transport coefficient.
The well known set of diagrams corresponding to the fluctuations corrections
to the longitudinal conductivity is presented in Fig.~\ref{fig:diagrams}. In
general all these diagrams may contribute to the leading correction to the
transverse conductivity, but actually this is not the case. It was shown in
Ref.~\onlinecite{Fukuyama1971} that the  anomalous Maki-Thompson
correction (illustrated in Fig.~\ref{fig:diagrams}a) is simply equal to $\delta\sigma_{xy}^{AMT}=-2\omega_{c}\tau\delta%
\sigma_{xx}^{AMT}(H,T)$.  Therefore, we do not have to dwell on the
derivation of this contribution. Furthermore, we obtained that out of the
remaining ten diagrams contributing to $\delta\sigma_{xx}$ only few give
non-zero contribution to $\delta\sigma_{xy}$. These are the Aslamazov-Larkin
term, Fig.~\ref{fig:diagrams}(k), and two of the density of state terms,
Fig.~\ref{fig:diagrams}(g) and ~\ref{fig:diagrams}(h). Although all other
diagrams have non-zero contribution to $\delta\sigma_{xy}$ when estimated
separately, their sum vanishes. In addition, we have discovered a new
contribution to the Hall current, which is presented in Fig.~\ref{fig:new1}.
The contribution of this term to $\sigma_{xx}$ is smaller by a factor of $%
T\tau$ than those from the set of ten diagrams in Fig.~\ref{fig:diagrams}. In contrast,
its contribution to the Hall conductivity is of the same order as the rest
of the terms.

The entire dependence on the magnetic field is incorporated through the
propagators of the quasiparticles, superconducting fluctuations and
Cooperons (which describe the multiple scattering of two quasiparticles by
impurities). Since we are interested in linear response to the electric
field, all propagators entering the diagrams are calculated at thermal
equilibrium. The equation for the quasiparticles Green's function at
equilibrium in the presence of a magnetic field is:
\begin{align}  \label{eq:3-G_eq}
&\left[\epsilon+\frac{1}{2m}\left(\boldsymbol{\nabla}-\frac{ie}{c}\mathbf{A}(%
\mathbf{r},t)\right)^2\hspace{-1.5mm} -\hspace{-0.5mm}V_{\text{imp}}(\mathbf{r})+\mu\phantom{\frac{1}{1}}\hspace{%
-2mm}\right]g^{R,A}(\mathbf{r},\mathbf{r}^{\prime};\epsilon)  \notag \\
&-\int{d\mathbf{r}_{\scriptscriptstyle1}}\Sigma_{eq}^{R,A}(\mathbf{r},%
\mathbf{r}_{\scriptscriptstyle1};\epsilon)g^{R,A}(\mathbf{r}_{%
\scriptscriptstyle1},\mathbf{r}^{\prime},\epsilon) =\delta(\mathbf{r}-%
\mathbf{r}^{\prime}),
\end{align}
Here, $\Sigma_{eq}$ is the quasiparticle self-energy at equilibrium. The
Green's function depends on the two spatial coordinates and not only on the
relative one due to the impurities potential, $V_{\text{imp}}(\mathbf{r})$, and the
vector potential $\mathbf{A}(\mathbf{r},t)$. The equilibrium Green's
function can be written as a product of the phase factor, $\exp\{ie\int_{%
\mathbf{r}^{\prime}}^{\mathbf{r}}\mathbf{A}(\mathbf{r}_{\scriptscriptstyle%
1})d\mathbf{r}_{\scriptscriptstyle1}/c\}$, and the gauge invariant Green's
function, $\hat{\tilde{G}}_{eq}$. In the presence of a uniform (and constant
in time) magnetic field, this representation of the Green's function takes
the following simple form:
\begin{align}  \label{eq:3-Phase}
\hat{g}(\mathbf{r},\mathbf{r}^{\prime};\epsilon)=\hat{\tilde{g}}(\mathbf{r},%
\mathbf{r}^{\prime};\epsilon)e^{-ie\mathbf{B}\cdot\left[(\mathbf{r-r}%
^{\prime})\times(\mathbf{r+r}^{\prime})\right]/4c}.
\end{align}
Then, the retarded and advanced components of $\hat{\tilde{g}}$ satisfy the
equation
\begin{align}  \label{eq:3-G_eqGaugeInv}\nonumber
&\left[\epsilon+\frac{1}{2m}\left(\boldsymbol{\nabla}-i\frac{e\mathbf{B}}{2c}%
\times(\mathbf{r}-\mathbf{r}^{\prime})\right)^2\hspace{-2mm}-V_{\text{imp}}(\mathbf{r})-\tilde{%
\Sigma}_{eq}^{R,A}+\mu\right] \\
&\times \tilde{g}^{R,A}(\mathbf{r},\mathbf{r}^{\prime};\epsilon)=\delta(%
\mathbf{r}-\mathbf{r}^{\prime}),  
\end{align}
where, the product of the Green's function and the self-energy should be
understood as a convolution in real space. Now the entire dependence of the
gauge invariant Green's function on the center of mass coordinate is due to
the impurities. After averaging over disorder, the gauge invariant part of
the Green's function $\hat{g}$ becomes translational invariant, i.e., it
is a function of the relative coordinate $\boldsymbol{\rho}=\mathbf{r}-%
\mathbf{r}^{\prime}$ alone (see Ref.~\onlinecite{Khodas2003} and references
therein):
\begin{align}  \label{eq:3-G_eqGaugeInv2}
&\left[\epsilon+\frac{1}{2m}\left(\frac{\partial^2}{\partial\boldsymbol{\rho}%
^2}-\frac{(e\mathbf{B}\times\boldsymbol{\rho})^2}{4c^2}\right)-\tilde{\Sigma}%
_{eq}^{R,A}+\mu\pm\frac{i}{2\tau}\right] \\
&\times \tilde{g}^{R,A}(\boldsymbol{\rho},\epsilon)=\delta(\boldsymbol{\rho}%
).  \notag
\end{align}
We restrict the calculation to the limit $\omega_c\tau\ll1$. Therefore, we
may neglect the dependence of $\tilde{G}$ on the magnetic field entering
through the Landau quantization of the quasiparticles states. Then, the only
dependence of the quasiparticle Green's functions on the magnetic field is
through the phase as described in Eq.~\ref{eq:3-Phase}. We wish to point out
that in the normal state, the permeability is close to unity and,
correspondingly, we do not distinguish between $B$ (the magnetic flux
density) and the magnetic field $H$.

Unlike the quasiparticles, the Landau quantization of the collective modes
(the fluctuations of the superconducting order parameter) cannot be
neglected when $\Omega_c/T>1$. The equilibrium propagator of the
superconducting fluctuations, like the quasiparticle Green's functions, can
be separated into the phase factor $\exp\{2ie\int_{\mathbf{r}^{\prime}}^{\mathbf{r}}\mathbf{A}(\mathbf{r}_{\scriptscriptstyle1})d\mathbf{r}_{%
\scriptscriptstyle1}/c\}$ and the gauge invariant part, $\tilde{L}$. The
gauge invariant part, $\tilde{L}$, can be written using the Landau level
quantization, $\tilde{L}^{R,A}(\mathbf{r},\mathbf{r}^{\prime};\omega)=\sum_{N}%
\varphi_{N,0}(\mathbf{r-r}^{\prime})\tilde{L}_{N}(\omega)$, where:
\begin{subequations}
\begin{align}  \label{eq:3-Lfinal}
&\tilde{L}_N^{R,A}(\omega) 
=-\frac{1}{\nu}\hspace{-1mm}\left[\ln\hspace{-1mm}%
\left(\frac{T}{T_c}\right)\hspace{-1mm}+\psi_{\scriptscriptstyle{R,A}%
}\left(\omega,N\right)\hspace{-1mm}-\hspace{-1mm}\psi\left(\frac{1}{2}\right)%
\hspace{-1mm}+\hspace{-1mm}\varsigma \omega\right]^{-1}\hspace{-3mm};  
\end{align}
\begin{align}  \label{eq:3-Digamma}
\psi_{\scriptscriptstyle{R,A}}\left(\omega,N\right)=\psi\left(\frac{1}{2}\mp%
\frac{{i}\omega}{4\pi{T}}+\frac{\Omega_c(N+1/2)}{4\pi{T}}\right).\hspace{-1mm%
}
\end{align}
\end{subequations}
Here, $\psi(x)$ is the digamma function, N is the index of the Landau level
and $\varphi_{N,n}(\mathbf{r})$ is the wave function of a particle in the N-th Landau level solved in the symmetric gauge.  As we have already
discussed, the appearance of the parameter $\varsigma$ in Eq.~\ref{eq:3-Lfinal} introduces
the particle-hole asymmetry into the propagator of the superconducting
fluctuations. In a similar way, the gauge invariant part of the Cooperon can be  written
in terms of the Landau levels:
\begin{align}  \label{eq:3-Cooperon}
\tilde{C}_N^{R,A}(\epsilon,\omega-\epsilon)=\frac{1}{\mp{i}%
(2\epsilon-\omega)\tau+\Omega_c\tau(N+1/2)},
\end{align}

In the derivation of the Aslamazov-Larkin, Fig.~\ref{fig:diagrams}(k), and
density of states diagrams, Figs~\ref{fig:diagrams}(g) and~\ref{fig:diagrams}%
(h), we can neglect the dependence of the quasiparticles on the magnetic
field. This is because the contributions from the phase associated with the
quasiparticle Green's functions (see Eq.~\ref{eq:3-Phase}) add to zero. \begin{widetext}Then
the integration over the quasiparticle degrees of freedom is trivial. As a
result, the Aslamazov-Larkin term becomes ($e<0$):
\begin{align}\label{eq:3-j_AL}
j_{AL}^{y}&=i\frac{e^2E_x}{8\pi^2}\nu^2\text{sign}(H)\int{d}\omega\sum_{N=0}^{\infty}(N+1)\frac{\partial{n_P(\omega)}}{\partial\omega}
\left[\psi_{R}(\omega,N)+\psi_{A}(\omega,N)-\psi_{R}(\omega,N+1)-\psi_{A}(\omega,N+1)
\right]\\\nonumber&\times
\left[\psi_{R}(\omega,N)-\psi_{R}(\omega,N+1)\right]
\left[\tilde{L}_{N}^R(\omega)\tilde{L}_{N+1}^A(\omega)-\tilde{L}_{N+1}^R(\omega)\tilde{L}_{N}^A(\omega)\right]
+i\frac{e^2E_y}{2\pi^2}\nu^2\text{sign}(H)\int{d}\omega\sum_{N=0}^{\infty}(N+1)n_P(\omega)\\\nonumber
&\times\left[\psi_{R}(\omega,N)-\psi_{R}(\omega,N+1)\right]^2
\left[\frac{\partial{\tilde{L}}_{N}^R(\omega)}{\partial\omega}\tilde{L}_{N+1}^R(\omega)-\frac{\partial{\tilde{L}}_{N+1}^R(\omega)}{\partial\omega}\tilde{L}_{N}^R(\omega)\right]+c.c.
\end{align}
and the density of states contribution is:
\begin{align}\label{eq:3-Hall3C}
&{j}_{DOS}^{y}=-\frac{e^2E_x}{4\pi^2}\nu\text{sign}(H)\int{{d}\omega}\sum_{N\geq0}(N+1)\left\{
\frac{-i}{2}\frac{\partial{n}_P(\omega)}{\partial\omega}\tilde{L}_{N}^{R}(\omega)
\left[\frac{\Omega_c(N+1)-\Omega_cN}{(4\pi{T})}\psi_{R}'(\omega,N)\right.\right.\\\nonumber
&\left.\left.+\psi_{R}(\omega,N)
-\psi_{R}(\omega,N+1)-
\frac{\Omega_c(N+1)-\Omega_cN}{4\pi{T}}\psi_{A}'(\omega,N)-
\psi_{A}(\omega,N)
+\psi_{A}(\omega,N+1)\right]\right.\\\nonumber
&\left.-\frac{n_P(\omega)}{4\pi{T}}\tilde{L}_{N}^{A}(\omega)\left[\frac{\Omega_c(N+1)-\Omega_cN}{4\pi{T}}\psi_{A}''(\omega,N)+\psi_{A}'(\omega,N)
-\psi_{A}'(\omega,N+1)\right] - (N\leftrightarrow{N+1})\right\} +c.c.
\end{align}
\end{widetext}
The notation $N\leftrightarrow{N+1}$ means that $N$ is replaced by $N+1$ and
the other way around in all the terms inside the curly brackets. In both terms some of the
propagators of the collective modes (the superconducting fluctuations and
Cooperons) are  functions of the $N$-th Landau level while the index for the
others propagators is $N+1$. This is due to the Lorentz force turning the
collective modes from the $x$ into the $y$ direction. For more
details of the derivation see Appendix~\ref{Derivation}. At low $H$ for which $%
\Omega_c\ll4\pi{T}$, the discrete sum over the Landau levels can be
replaced by an integral (the continuum limit).

In contrast to the Aslamazov-Larkin and the density of states corrections,
in the derivation of the new contribution illustrated in Fig.~\ref{fig:new1}
the Lorentz force acts on the quasiparticle in order to turn the current.
Hence, we cannot ignore the magnetic field entering their phase.
Consequently, the integration over the quasiparticle degrees of freedom is more subtle than in the derivation of the previous terms, see
Appendix~\ref{Derivation} for details. The result of integrating out the quasiparticles is:
\begin{align}  \label{eq:New}
&{j}_{new}^{y}=-i\frac{e^2E_x}{32\pi^2}\nu\Omega_c^2\left[\frac{1}{%
\varepsilon_F}+\frac{\nu^{\prime}(\mu)}{\nu(\mu)}\right] \text{sign}(H) \\
&\times\int{d\omega}\sum_{N\geq0} \left\{ \left(\frac{1}{4\pi{T}}%
\right)^2n_P(\omega)\tilde{L}_{N}^{A}(\omega)\psi^{\prime\prime}_{A}(%
\omega,N)\right.  \notag \\
&\left.+\frac{i}{8\pi{T}}\frac{\partial{n_P(\omega)}}{\partial\omega}\tilde{L%
}_{N}^{R}(\omega)\left[\psi_{R}^{\prime}(\omega,N)-\psi_{A}^{\prime}(%
\omega,N) \right] \right\}+c.c.  \notag
\end{align}
In the above expression all collective mode propagators have the same Landau level index. Although it is not evident, this contribution is proportional to the cyclotron frequency of the quasiparticles. Comparison with the correction to the longitudinal conductivity
arising from the modification of the tunneling density of states by the fluctuations,~\cite{Konstantin Georg,KamenevLevch} shows that the new term describes how the tunneling density of states reveals itself in the transverse conductivity.

\section{Fluctuations corrections to the Hall effect}

\label{sec:phasediagram}
\begin{figure}[]
\begin{flushright}\begin{minipage}{0.5\textwidth}  \centering
      \includegraphics[width=0.95\textwidth]{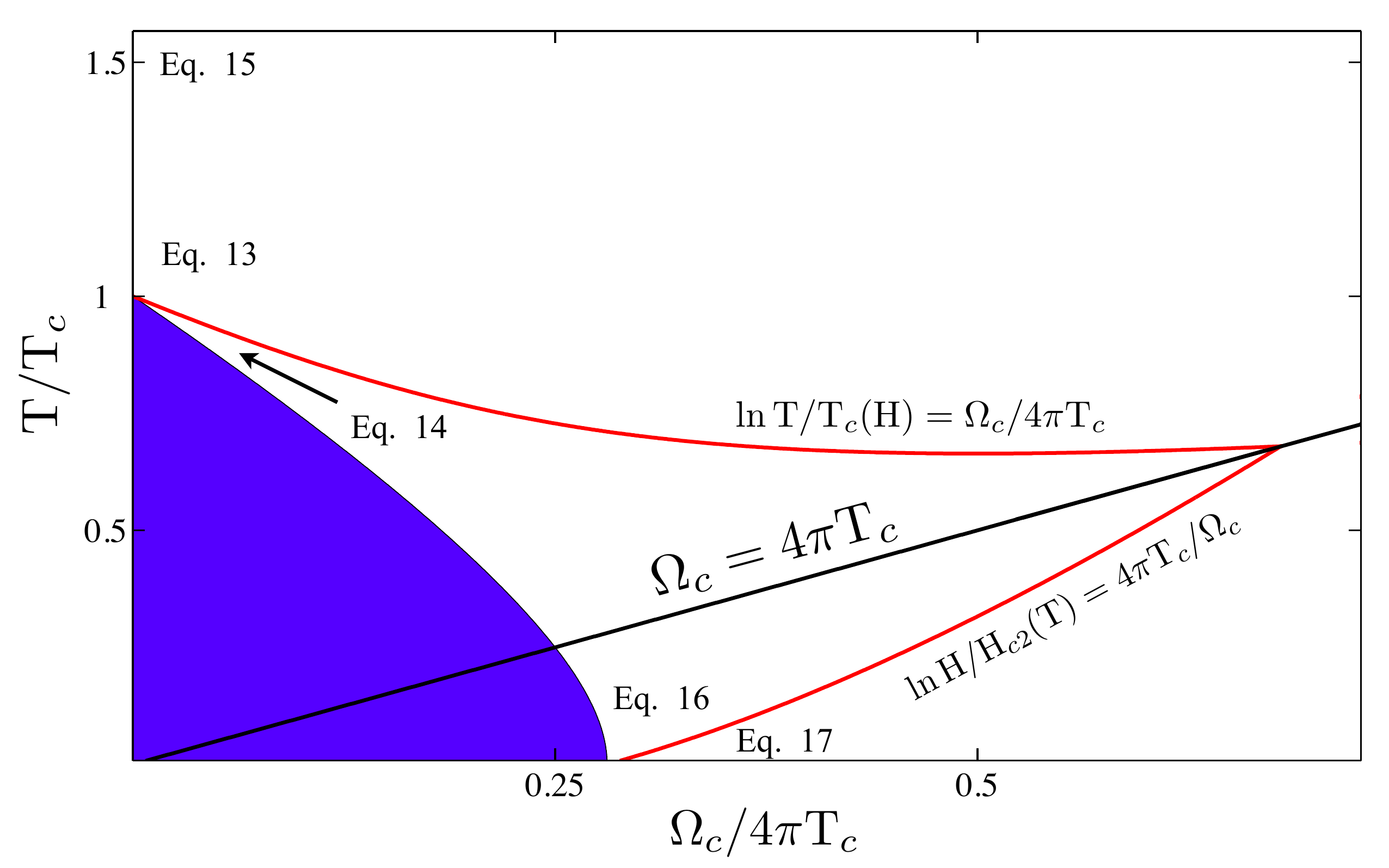} \hspace{0.05in}
               \caption[0.4\textwidth]{\small The phase diagram for the corrections to the Hall conductivity $\delta\sigma_{xy}$. The equations indicated on the phase diagram correspond to the expressions for $\delta\sigma_{xy}$ written in the text. $\Omega_c=4eHD/c$ is the cyclotron frequency corresponding to the superconducting fluctuations  in the diffusive regime.} \label{fig:PhaseDiagramHall}
\end{minipage}\end{flushright}
\end{figure}

We now present  the leading corrections to the Hall resistance in the
different regions of the phase diagram plotted in Fig.~\ref {fig:PhaseDiagramHall}. A similar phase diagram has been previously discussed
in a study of the Nernst Effect in amorphous superconducting films.~\cite{NernstLetter,Nernst} As shown in Fig.~\ref {fig:PhaseDiagramHall}
the phase diagram is divided into many subregions. This is because the magnetic field plays a double role; not
only does it drive the transition between the metallic normal state and the
superconducting one, it also quantizes the collective modes in the Cooper
channel (both the superconducting fluctuations and Cooperons). The shaded area corresponds to the superconducting phase which is bounded by  the line $T=T_c(H)$. There are two crossover lines in the vicinity of the transition. In the area below the line $\ln{T/T_c(H)}=\Omega_c/4\pi{T}$ the Landau Level quantization of the superconducting fluctuations becomes essential. The other line, $ln{H/H_{c2}(T)}=4\pi{T}/\Omega_c$, separates the regions of classical and quantum fluctuations at low temperatures. The low-$H$ and high-$T$ region is separted from the high-$H$ and low-$T$ region by the line  $\Omega_c=4\pi{T}$.

As we explained in the previous section, different contributions to the Hall conductivity are characterized by the way  the magnetic field deflects the current to the transverse direction. The magnetic field can turn the current via the collective modes or the quasiparticles. The first case yields contributions proportional to  $\varsigma\Omega_c\sim\omega_c\tau/\lambda$, where $\lambda$ is the dimensionless coupling constant of the attractive electron-electron interaction in the Cooper channel. The other possibility results in corrections that do not  contain the large factor $1/\lambda$.

\begin{figure}[tp]
\begin{flushright}\begin{minipage}{0.5\textwidth}  \centering
      \includegraphics[width=0.95\textwidth]{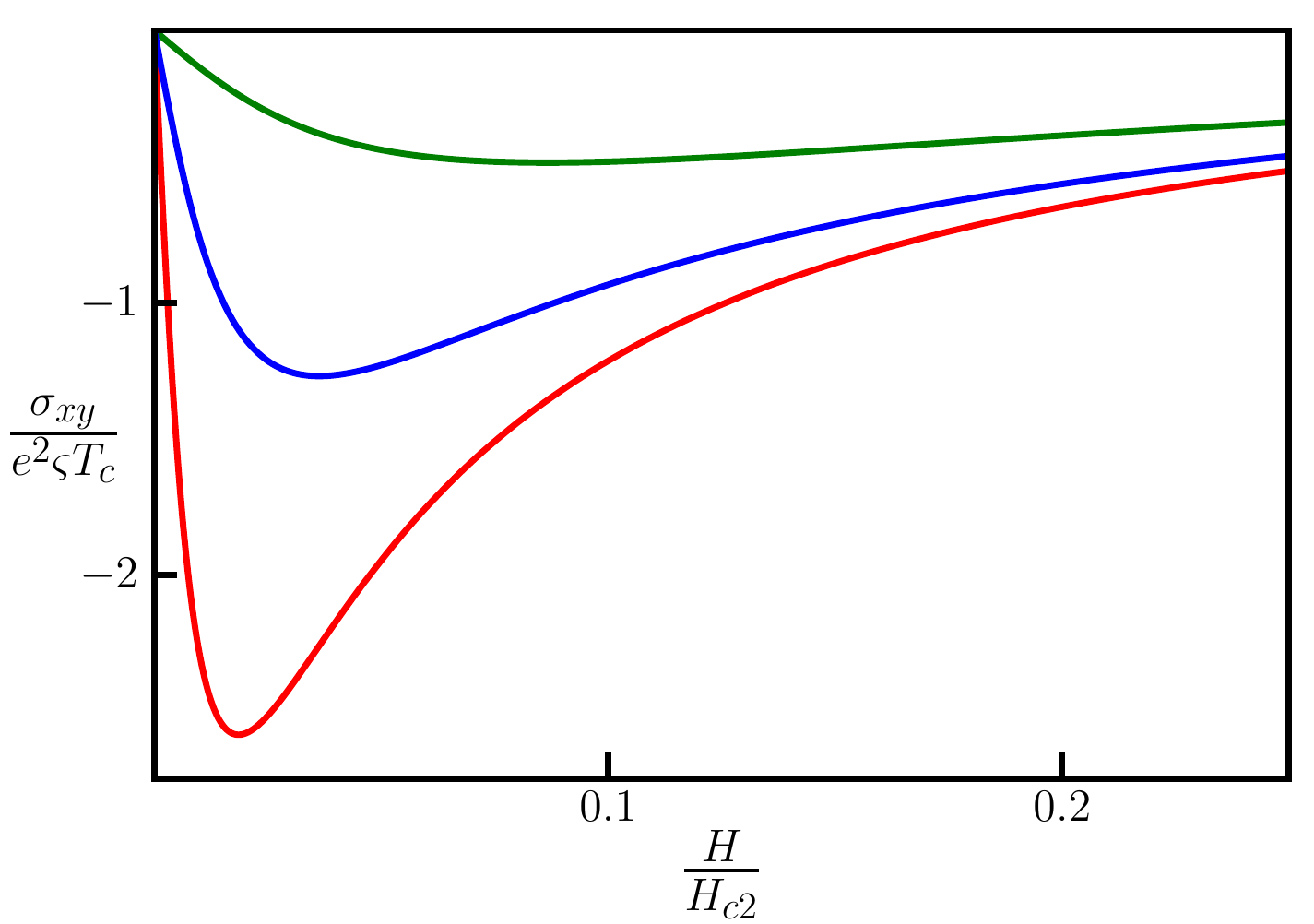} \hspace{0.05in}
               \caption[0.4\textwidth]{\small Corrections
                to the Hall conductivity $\delta\sigma_{xy}$ as described by  Eq.~\ref{eq:ALTc} for $T=1.01T_c$ (red curve), $T=1.02T_c$ (blue curve) and $T=1.05T_c$ (green curve). The Hall conductivity is given in units of $e^2|\varsigma|T_c$ 
               }\label{fig:sigmaplot}
\end{minipage}\end{flushright}
\end{figure}

Close to the line of phase transition, $T\gtrsim{T_c(H)}$, and for a small
magnetic field, $\Omega_c\ll4\pi{T}$, the leading correction to $\sigma_{xy}$ is given by the Aslamazov-Larkin term:
\begin{equation}
\label{eq:ALTc}
\delta\sigma_{xy}=\frac{2e^2\varsigma T\nu}{\pi}\text{sign}(H)\sum_{n}\left(  n+1\right)
\frac{\left[\tilde{L}_{n}\left(0\right)-\tilde{L}_{n+1}\left(0\right)  \right]  ^{3}%
}{\left[  \tilde{L}_{n+1}\left(0\right)+\tilde{L}_{n}\left(  0\right)  \right]  ^{2}}.
\end{equation}
The above equation is derived from Eq.~\ref{eq:3-j_AL} by expanding to the first order in $\varsigma T$.  In addition, we integrated over the frequency $\omega$ only up to $T$ (accounting  for the classical fluctuations alone). This correction to the Hall conductivity, just like the Drude term is negative, because $\varsigma<0$.  Note that here, and in what follows, we consider negative charge carriers $e<0$. As we show in  Fig.~\ref{fig:sigmaplot}, for $T>T_c(H=0)$,  the correction to the Hall conductivity is  a non-monotonic function of the magnetic field. In the close vicinity of $T_c(H=0)$, $\delta\sigma_{xy}$ has a peak at $H^{*}=1.3\frac{\phi _{0}}{2\pi } \frac{\ln T/T_{c}}{\xi^{2}}$ which up to a factor of $1.3$ coincides with the ghost field observed in measurements of the Nernst effect~\cite{Pourret2007} (here $\xi^2=\pi{D}/8T_c$). The above expression has been successfully used to fit the data obtained in  recent measurements of the Hall conductivity in amorphous Tantalum Nitrade films (see Fig.~5 in Ref~\onlinecite{Kapitulnik}).

As the magnetic field goes to zero and $T>T_c(H=0)$, the discrete sum over the Landau levels can be replaced by a continuous integral. Then the correction to the Hall conductivity from Eq.~\ref{eq:ALTc} becomes:
\begin{equation}
\label{GLInt}
\delta\sigma_{xy}=e^2\frac{\varsigma\Omega_{c}}{96}\text{sign}(H)\left(  \frac{1}{\ln{T/T_{c}(H)}  }\right)  ^{2}.
\end{equation}
Curiously, close to the  transition  the divergence of the Hall conductivity, $\delta\sigma_{xy}\sim1/\ln^2({T}/T_c)$, is stronger than the one known for the longitudinal conductivity,~\cite{Aslamazov1968} $\delta\sigma_{xx} \sim1/\ln({T}/T_c)$.  When $T<T_c(H=0)$, the Landau level quantization is essential. Moreover, far below the line $\ln{T/T_c(H)}=\Omega_c/4\pi{T}$ only the lowest Landau level contributes to the sum, and one gets:
\begin{equation}
\label{GL0}
\delta\sigma_{xy}=\frac{2e^2\varsigma T_{c}}{\pi}\text{sign}(H)\frac{1}{\ln{T/T_{c}(H)}}.
\end{equation}
Note that this expression does not contain the magnetic field as a prefactor.

At $T\gg T_c$ but still at a small magnetic
field, the process described by the new contribution  introduced in this paper (see Fig.~\ref{fig:new1}) dominates:
\begin{align}  \label{eq:3-farfromTc}
&\delta\sigma_{xy}\approx\frac{e^2{\omega_c\tau}}{4\pi^2}\text{sign}{H}\ln\left(\frac{\ln{1/T\tau}}{\ln{T/T_c}}\right).
\end{align}
The new term, and therefore also the leading correction to the Hall conductivity at $T\gg{T}_c$, is proportional to $\omega_c$, because in this case the Lorentz force turning the current from the longitudinal to the transverse direction acts on the quasiparticles rather than the superconducting fluctuations. Comparing Eq.~(\ref{eq:3-farfromTc}) with the correction to the longitudinal conductivity in this region~\cite{Alt83}, one may observe that $\delta\sigma_{xy}=-\frac{\omega_c\tau}{2}\delta\sigma_{xx}$.

In the vicinity of the magnetic field driven quantum critical point, $H\approx{H}_{c2}(T=0)$, all three terms discussed in the previous section as well as the anomalous Maki-Thompson term contribute comparably to the Hall
conductivity.
In the classical regime where $\ln{H}/H_{c2}(T)<4\pi{T}/\Omega_c\ll1$ the Hall conductivity is
\begin{align}  \label{eq:3-Thermal}
&\delta\sigma_{xy}\approx\frac{2e^2}{\pi\ln{H/H_{c2}}}\text{sign}{H}%
\left(\varsigma {T}-\frac{21T}{8\varepsilon_F}\right).
\end{align}
Here the Hall conductivity depends on the magnetic field only via $\ln{H}/H_{c2}$ which measures the distance to the phase transition. In the quantum regime, $4\pi{T}/\Omega_c<\ln{H}/H_{c2}(T)\ll1$, the Hall conductivity acquires the form:
\begin{align}  \label{eq:3-Quantum}
&\delta\sigma_{xy}\approx\frac{e^2\text{sign}{H}}{2\pi^2}\left({\omega_c\tau}%
-\frac{2\varsigma \Omega_c}{3}\right)\ln{\frac{1}{\ln{H/H_{c2}}}}.
\end{align}

Finally, we wish to emphasize how the Landau quantization of the collective modes enters the Hall conductivity. In general,  to obtain the fluctuation corrections to $\sigma_{xy}$ one must sum over all Landau levels. However,  there are limiting cases in which the sum can be simplified: (i) $H\rightarrow0$ and (ii) $\ln{T}/T_{c}(H)\ll\Omega_c/4\pi{T}$. In the first case, the sum over $N$ can be replaced by an integral. This simplification has been used to obtain Eqs.~\ref{GLInt} and~\ref{eq:3-farfromTc}. In the second case, the critical behavior is determined by the contribution from the lowest Landau level. Consequently, in deriving Eqs.~\ref{GL0},~\ref{eq:3-Thermal}, and~\ref{eq:3-Quantum} we neglected terms with $N>0$.

In conclusion we extended the previous calculations of the Hall conductivity~\cite{Fukuyama1971,Larkin1995} to a broader range of temperatures and
magnetic fields. The fluctuations corrections can be divided into two groups. The first contains terms proportional to $\varsigma\Omega_c$ and includes the Aslamazov Larkin contribution (Fig.~\ref{fig:diagrams}k) and part of the density of state corrections (Figs.~\ref{fig:diagrams}g and~\ref{fig:diagrams}h) The other group includes the  new contribution $\delta\sigma_{xy}^{NEW}$  (Fig.~\ref{fig:new1}) that was not considered before, and the anomalous Maki-Thompson term (Fig.~\ref{fig:diagrams}a). These corrections are proportional to $\omega_c\tau$. Unlike the anomalous Maki-Thompson correction, the new contribution modifies the Hall resistivity. This becomes obvious if we rewrite the Hall resistivity in terms of the two components of the conductivity tensor, $\rho_{xy}=-\sigma_{xy}/(\sigma_{xx}^2+\sigma_{xy}^2)\approx-\sigma_{xy}/\sigma_{xx}^2$, and  extract the fluctuation correction to the resistivity, $\delta\rho_{xy}=-\delta\sigma_{xy}/\sigma_{xx}^2+2\sigma_{xy}\delta\sigma_{xx}/\sigma_{xx}^3$, with $\sigma_{xy}=-\omega_c\tau\sigma_{xx}$. Since $\delta\sigma_{xy}^{AMT}=-2\omega_c\tau\delta\sigma_{xx}^{AMT}$, the anomalous Maki-Thompson correction to $\rho_{xy}$ vanishes, while the correction from $\delta\sigma_{xy}^{NEW}$ remains. Our results for the different regimes of the phase diagram are  summarized in Fig.~\ref{fig:PhaseDiagramHall}.

\textit{Acknowledgement:} This work is supported by Pappalardo Fellowship (KM) and the National Science Foundation grant  NSF-DMR-1006752 (AMF). KT and AMF are supported by NHRAP. We would like to thank G. Schwiete for helpful discussions.

\appendix

\section{Particle-hole asymmetry and  superconducting fluctuations}
\label{sec:asymmetry}
Here we will explain the mechanism of appearance  of the parameter $\varsigma$ in the propagator of
superconducting fluctuations given in Eq.~\ref{eq:3-Lfinal}.
For that we  calculate $\hat{L}$ taking into account the
dependence of the density of states and velocity of the quasiparticles on energy. In the normal state, the
quasiparticles are described in terms of the Fermi liquid theory where the standard
approximation is to consider   the density of states and velocity in the vicinity of the Fermi energy as constants.
The dependence of the Fermi liquid parameters on energy leads only to small corrections and can be usually ignored. However, under this approximation the propagator of
superconducting fluctuations satisfies $L^{R}\left( \omega \right) =L^{A}\left( -\omega \right)$ and, consequently,
the fluctuations corrections to the Hall effect vanish. Therefore, when studying the Hall effect, we have to go
beyond the Fermi liquid approximation. Note that although the fluctuations in
superconducting films are effectively two-dimensional,
the quasiparticles in a not too thin film are still three-dimensional and, hence,
the density of states $\nu$ is not a constant.

The propagator of superconducting fluctuations at equilibrium satisfies the
following equation:
\begin{align}
L^{R,A}
(\mathbf{r},t;\mathbf{r}^{\prime},t^{\prime})=\frac{1}{\nu_0}\left(-\lambda^{-1}+\Pi^{R,A}(%
\mathbf{r},t;\mathbf{r}^{\prime},t^{\prime})\right)^{-1}.
\end{align}
In this work we study effects of superconducting fluctuations in the gaussian  approximation. After averaging over disorder,
the polarization operator  can be written in terms of the Cooperon
and the quasiparticle Green's functions:
\begin{align}
&\hat{\Pi}(\mathbf{r},t;\mathbf{r}^{\prime},t^{\prime})\\\nonumber&=\frac{1}{\nu_0}\int{d\mathbf{r}_1}%
dt_1\hat{g}(\mathbf{r},t;\mathbf{r}_1,t_1)\hat{g}(\mathbf{r},t;\mathbf{r}%
_1,t_1)\hat{C}(\mathbf{r}_1,t_1;\mathbf{r}^{\prime},t^{\prime}).
\end{align}
It will be enough to find $\Pi$ in the absence of magnetic field,
and reintroduce the magnetic field in the end. Then, the  calculation
can be done in momentum and frequency space, and the Cooperon becomes:
\begin{align}  \label{eq:CooperonApp}
&C^{R}(\mathbf{q},\epsilon, \omega-\epsilon)\\\nonumber&=\left[1-V_{\text{imp}}^2\int\frac{d%
\mathbf{k}}{(2\pi)^3} g^{R}(\mathbf{k},\epsilon)g^{A}(\mathbf{q-k}%
,\omega-\epsilon)\right]^{-1}.
\end{align}
The particle-hole asymmetry enters  the calculation of the Cooperon in numerous ways. First of all, the non-constant density of states affects  the elastic scattering time, and hence, modifies the quasiparticle Green's function:
\begin{align}
g^{R,A}(\mathbf{k},\epsilon)=\left[\epsilon-\xi_{\mathbf{k}}\pm{i}\pi{V}_{\text{imp}}^2\nu(\epsilon)\right]^{-1}.
\end{align}
For a parabolic spectrum of three-dimensional quasiparticles, $\nu \left( \epsilon \right) \approx \nu _{0}\left( 1+\epsilon /2\varepsilon _{F}\right)$.
Similarly,  the integration over the momentum in Eq.~\ref{eq:CooperonApp}, is sensitive to the energy dependence of the density of states and velocity. In practice, however, the analysis of the leading contribution shows that only the modification of the quasiparticle Green's functions are important. Then, expanding the density of states in the Green's functions, one gets:
\begin{align}
C^{R,A}(\mathbf{q},\epsilon,\omega-\epsilon)=\frac{1+\omega/4\varepsilon_F}{\mp%
{i}(2\epsilon-\omega)\tau+Dq^2\tau},
\end{align}
where $\tau=(2\pi{V}_{\text{imp}}^2\nu_0)^{-1}$ is the elastic scattering time at the Fermi energy calculated in the Born approximation.

We can see that the particle-hole asymmetry modifies the Cooperon by the factor $(1+\omega/4\varepsilon_F)$. Correspondingly, the polarization
operator becomes:
\begin{widetext}
\begin{align}
\Pi^{R,A}(\mathbf{q},\omega)=-\left(1+\frac{\omega}{4\varepsilon_F}\right)\left[\psi\left(\frac{1}{2}+\frac{\mp{i}\omega+Dq^2}{4\pi{T}}\right)-\psi\left(\frac{1}{2}\right)+\ln\frac{T}{T_c}-\frac{1}{\lambda}\right].
\end{align}
Not too far from the superconducting transition, e.g., when $T\gtrsim T_{c}$, we can write the propagator $L^{R,A}(\mathbf{q},\omega)$ to the leading corrections due to the particle-hole asymmetry as:
\begin{align}
L^{R,A}(\mathbf{q},\omega)&=-\frac{1}{\nu _{0}}\left\{ \frac{1}{\lambda }+\left( 1+\frac{\omega }{%
4\varepsilon }\right) \left[ \ln \frac{T}{T_{c}}+\psi \left( \frac{1}{2}+\frac{\mp i\omega +Dq^{2}}{4\pi T}\right) -\psi \left( \frac{1}{2}\right) -\frac{1}{\lambda }\right] \right\} ^{-1}\\\nonumber
&\approx\frac{-1}{\nu_0}
\left[\ln\frac{T}{T_c}+\psi\left(\frac{1}{2}+\frac{\mp{i}\omega+Dq^2}{4\pi{T}}\right)-\psi\left(\frac{1}{2}\right)
-\frac{\omega}{4\varepsilon_F\lambda}\right]^{-1}.
\end{align}
\end{widetext}
Defining  $\varsigma=-1/4\varepsilon_F\lambda$, we get the expression for the propagator of the superconducting fluctuations used in the main text (see Eq.~\ref{eq:3-Lfinal}).  The asymmetry parameter $\varsigma$ can be rewritten as $\varsigma=-0.5d\ln{T_c}/d\ln{\mu}$, in accordance with Ref~\onlinecite{Larkin1995}. Furthermore, in the presence of magnetic field, the term $Dq^2$ in the propagator $L$ (as well as in the Cooperon) is quantized into the Landau levels, $Dq^2\rightarrow\Omega_c(N+1/2)$.  One may still use the obtained value for the parameter $\varsigma$ in the propagator $L$ as given in Eq.~\ref{eq:3-Lfinal} for the analysis of fluctuation effects in the Hall conductivity in the whole region $T$-$H$ of the superconducting transition, $T=T_c(H)$.

Finally, let us remark that although the asymmetry affects also the Cooperon, in
the derivation of the corrections to the Hall conductivity we neglected it.
Including the dependence of the Cooperon on the particle-hole asymmetry
leads to corrections which are smaller by a factor $T\tau\ll1$ or $1/\varepsilon_F\tau\ll1$
than the terms discussed in this paper.

\section{Derivation of the Hall conductivity}

\label{Derivation}

We apply here the quantum kinetic technique as described in Refs.~%
\onlinecite{NernstLetter,Nernst,QKE}. In the presence of superconducting
fluctuations we describe the system using two fields: the quasiparticle
field and the fluctuations of the superconducting order parameter. The
matrix functions $\hat{G}(\mathbf{r},\mathbf{r}^{\prime},\epsilon)$ and $%
\hat{\mathcal{L}}(\mathbf{r},\mathbf{r}^{\prime},\omega)$ written in the
Keldysh form~\cite{Keldysh1964,Rammer1986,Haug},
\begin{align}
\hat{F}(\mathbf{r},t;\mathbf{r}^{\prime},t^{\prime})=\left(
\begin{array}{cc}
F^{R}(\mathbf{r},t;\mathbf{r}^{\prime},t^{\prime}) & F^{K}(\mathbf{r},t;%
\mathbf{r}^{\prime},t^{\prime}) \\
0 & F^{A}(\mathbf{r},t;\mathbf{r}^{\prime},t^{\prime})%
\end{array}
\right),
\end{align}
(where $F$ can be either $G$ or $\mathcal{L}$) describe the propagation of
these two fields, respectively. The Keldysh components of the propagators
correspond to the generalized distribution functions. According to the
quantum kinetic approach the current can be written in terms of the
generalized distribution functions. For this purpose, we express the charge
density in terms of the propagators of the quasiparticles, $\hat{G}$, and
superconducting fluctuations, $\hat{\mathcal{L}}$. Since both the
quasiparticles and the superconducting fluctuations carry charge, they both
enter the continuity equation. Extracting the electric current from the
continuity equation we get:
\begin{align}  \label{eq:Current-Je}
&\mathbf{j}_{e}^{con}(\mathbf{r},t)=ie\hspace{-1mm}\int{d\mathbf{r}%
^{\prime}dt^{\prime}}\left[\mathbf{\hat{v}}(\mathbf{r},t;\mathbf{r}%
^{\prime},t^{\prime})\hat{G}(\mathbf{r}^{\prime},t^{\prime};\mathbf{r},t)%
\right]^{<} \\
&+ie\int{d\mathbf{r}^{\prime}dt^{\prime}}\hspace{-1mm}\left[%
\boldsymbol{\hat{\mathcal{V}}}(\mathbf{r},t;\mathbf{r}^{\prime},t^{\prime})%
\hat{\mathcal{L}}(\mathbf{r}^{\prime},t^{\prime};\mathbf{r},t)\right]%
^{<}+h.c.  \notag
\end{align}
Each of the terms in the current is a product of the renormalized velocity
and propagator. The matrix $\mathbf{\hat{v}}(\mathbf{r},t;\mathbf{r}%
^{\prime},t^{\prime})$ is the velocity of the quasi-particles renormalized
by the self-energy $\hat{\Sigma}(\mathbf{r},t;\mathbf{r}^{\prime},t^{\prime})
$:
\begin{align}  \label{eq:VelocityQP}
\mathbf{\hat{v}}(\mathbf{r},t;\mathbf{r}^{\prime},t^{\prime})&=-\frac{i}{2m}%
\left(\boldsymbol{\nabla}-\frac{ie}{c}\mathbf{A}(\mathbf{r})-%
\boldsymbol{\nabla}^{\prime}-\frac{ie}{c}\mathbf{A}(\mathbf{r}%
^{\prime})\right) \\
&\times\delta(\mathbf{r-r^{\prime}})\delta(t-t^{\prime}) -i(\mathbf{r-r}%
^{\prime})\hat{\Sigma}(\mathbf{r},t;\mathbf{r}^{\prime},t^{\prime}),  \notag
\end{align}
where $\mathbf{A}(\mathbf{r})$ is the vector potential. Similarly, we define
$\hat{\boldsymbol{\mathcal{V}}}(\mathbf{r},t;\mathbf{r}^{\prime},t^{%
\prime})=-i(\mathbf{r-r^{\prime}})\hat{\Pi}(\mathbf{r},t;\mathbf{r}%
^{\prime},t^{\prime})$ to be the "renormalized velocity" of the
superconducting fluctuations. Here $\hat{\Pi}$ is the polarization operator
in the Cooper channel (note that in fact $\hat{\boldsymbol{\mathcal{V}}}$
does not have the dimension of velocity). In general, all quantities in the
equation for the current depend on the external electric and magnetic fields.

Next, we derive the kinetic equation for the two propagators in the presence
of electric and magnetic fields. We consider linear response to the electric
field while keeping the entire dependence on the magnetic field. Then the $%
\mathbf{E}$-dependent quasiparticle Green's function is:
\begin{align}  \label{eq:G_TransInv}
&\hat{G}_{\mathbf{E}}(\mathbf{r},\mathbf{r}^{\prime},\epsilon)=\hat{g}%
\left(\epsilon\right) \hat{\Sigma}_{\mathbf{E}}\left(\epsilon\right)\hat{g}%
\left(\epsilon\right) \\
&-\frac{ie\mathbf{E}}{2}\left[\frac{\partial\hat{g}\left(\epsilon\right)}{%
\partial\epsilon}\mathbf{\hat{v}}_{eq}(\epsilon)\hat{g}\left(\epsilon\right)
-\hat{g}\left(\epsilon\right)\mathbf{\hat{v}}_{eq}(\epsilon)\frac{\partial%
\hat{g}\left(\epsilon\right)}{\partial\epsilon} \right].  \notag
\end{align}
The product of matrices should be understood as a convolution of the spatial
coordinates. In addition, we used the fact that we are interested in the
stationary solution for the Green's function in the presence of a DC
electric field. Hence, all Green's functions and self-energies are function
of the time difference $t-t^{\prime}$, and it was possible to Fourier
transform the equation from the relative time coordinate to the frequency $%
\epsilon$. In the above equation $\hat{g}$ is the equilibrium Green's
functions and $\hat{v}_{eq}$ is the quasiparticle velocity at equilibrium.
Note that the equation for the field dependent Green's function is a
self-consistent equation as it contains the $\mathbf{E}$-dependent
self-energy which is itself a function of $\hat{G}_{\mathbf{E}}$. In
addition, $\Sigma_{\mathbf{E}}$ may depend on the electric field through the
propagator of the superconducting fluctuations. The equation for the $%
\mathbf{E}$-dependent part of $\hat{\mathcal{L}}$ takes a form similar to
Eq.~\ref{eq:G_TransInv} for $\hat{G}_{\mathbf{E}}$:
\begin{align}  \label{eq:3-L_E}
&\hat{L}_{\mathbf{E}}(\omega) =-\hat{L}\hat{\Pi}_{\mathbf{E}}\hat{L} \\
&+ie\mathbf{E}\left[\frac{\partial\hat{L}(\omega)}{\partial\omega}\hat{%
\mathcal{V}}_{eq}(\omega)\hat{L}(\omega) -\hat{L}(\omega)\hat{\mathcal{V}}%
_{eq}(\omega)\frac{\partial\hat{L}(\omega)}{\partial\omega}\right].  \notag
\end{align}
Here $\hat{\mathcal{V}}_{eq}$ is the velocity of the superconducting
fluctuations at equilibrium, and $\Pi_{\mathbf{E}}$ is the electric field dependent polarization operator which depends on $G_{\mathbf{E}}$. The discussion of the equilibrium propagators $\hat{g}$ and $\hat{L}$ appears in the main text. In the following, we neglect the particle-hole asymmetry in $\hat{\mathcal{V}}_{eq}$ as well as in all the terms except $\hat{\mathcal{L}}$ since they result in less singular contributions than those discussed here.

\begin{figure}[pb]
\begin{flushright}\begin{minipage}{0.5\textwidth}  \centering
      \includegraphics[width=0.7\textwidth]{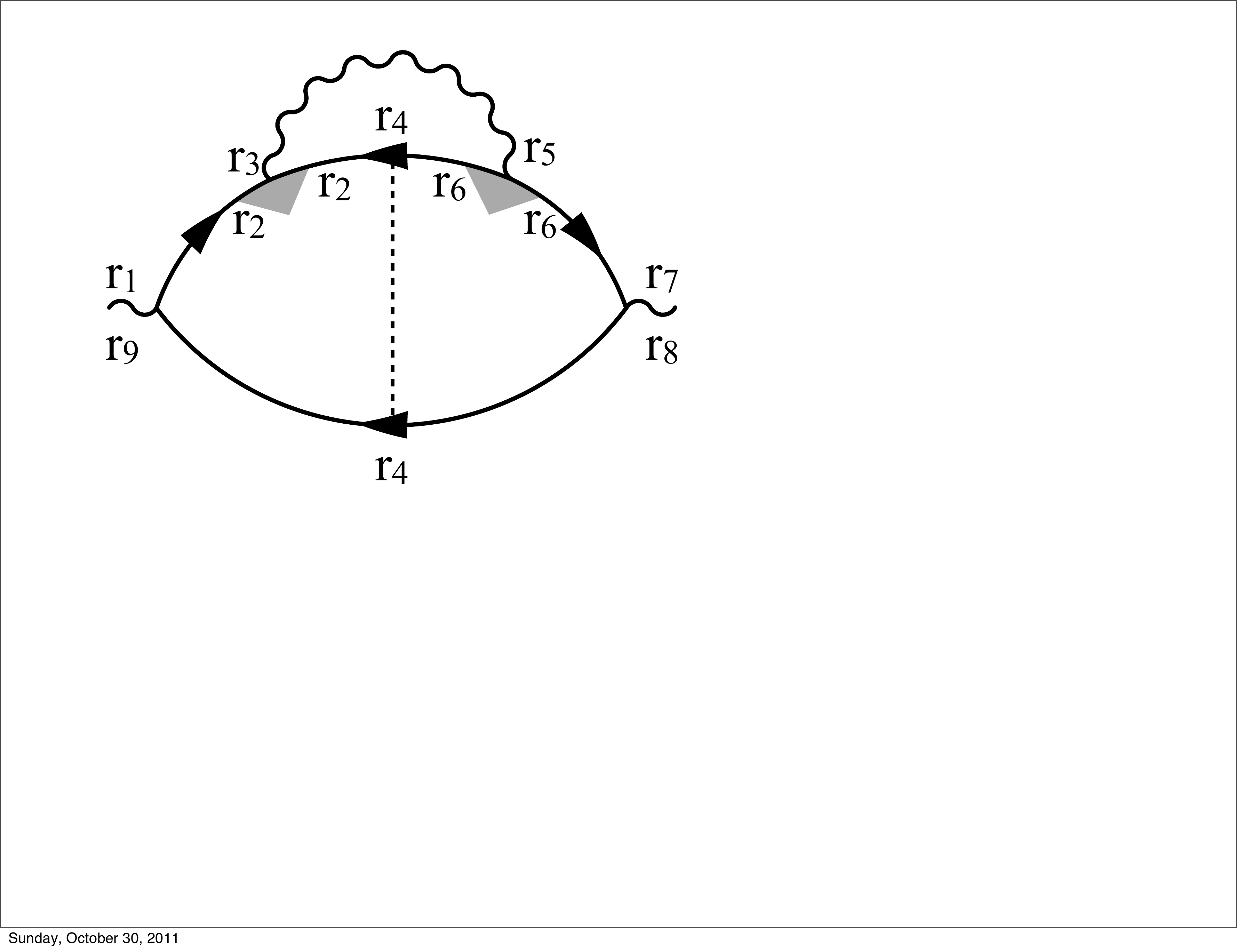} \hspace{0.05in}
               \caption[0.4\textwidth]{\small The new contribution to the Hall conductivity.} \label{fig:new2}
\end{minipage}\end{flushright}
\end{figure}

\begin{figure}[]
\begin{flushright}\begin{minipage}{0.5\textwidth}  \centering
      \includegraphics[width=0.75\textwidth]{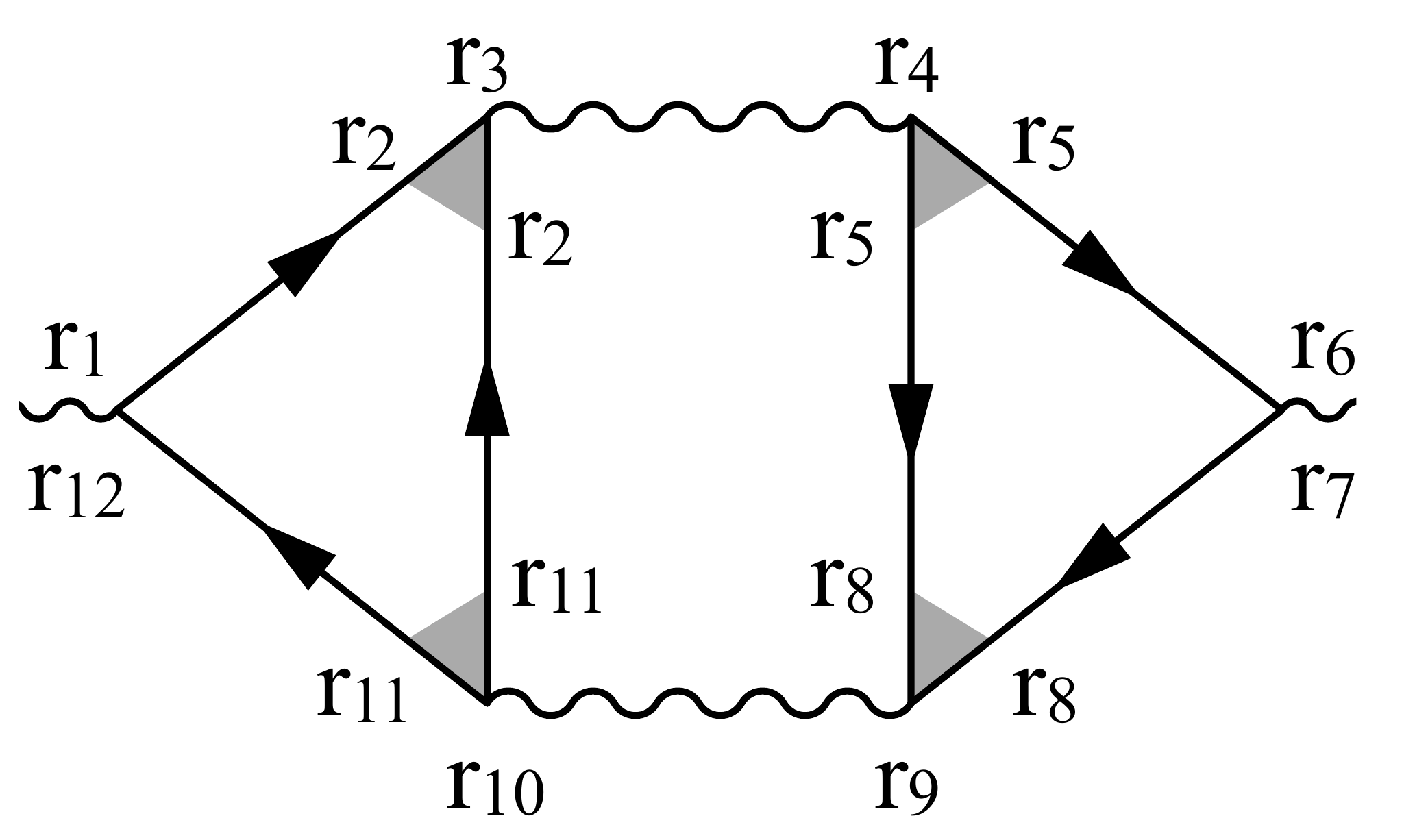} \hspace{0.05in}
               \caption[0.4\textwidth]{\small The Aslamazov-Larkin correction.} \label{fig:AL}
\end{minipage}\end{flushright}
\end{figure}

The next step in the derivation of the current is to insert the expression
for the $\mathbf{E}$-dependent propagators and velocities into Eq.~\ref%
{eq:Current-Je}. Up to now we only made two assumptions: (i) we restrict the
calculation to the regime of linear response to the electric field, (ii) we
consider classically weak magnetic field for which the cyclotron frequency
of the quasiparticles satisfy $\omega_c\tau\ll1$. As we are interested in
the Gaussian fluctuations, we will make further simplification by expanding
with respect to the superconducting fluctuations. Below we give a
diagrammatic interpretation for the dominant contributions to the Hall
conductivity. The expression for the vertices and the analytical structure
of these diagrams have been found from the quantum kinetic equation. The
quantum kinetic approach provides a simple and clear derivation of the Hall
conductivity, however, one can reach the same result using the standard Kubo
formula.

As we already explained, we can classify the contributions to the Hall
conductivity according to the way the current is deflected by the Lorentz
force. The first group containing the anomalous Maki-Thompson and the new
contribution includes terms in which the quasiparticles are used in order to
turn the current, while the current in the second group (Figs.~\ref%
{fig:diagrams}(g), \ref{fig:diagrams}(h), and~\ref{fig:diagrams}(k)) is
deflected using the collective modes. Let us first use, as an example, the
new term presented in Fig.~\ref{fig:new2}, to demonstrate how the magnetic
field enters these kind of contributions. Decomposing all propagators in the
diagram shown in Fig.~\ref{fig:new2} into the phase and gauge invariant
parts (see Eqs.~\ref{eq:3-Phase}, \ref{eq:3-Lfinal} and~\ref{eq:3-Cooperon}%
), we get:
\begin{widetext}
\begin{align}\label{eq:3-Hall2C2}
&{j}_{New}^{y}=-i\frac{e^2E_x}{4\pi\nu\tau\sqrt{2\pi\ell_H^2}}\int\frac{d\epsilon{d}\omega}{(2\pi)^2}d\mathbf{r}_2...d\mathbf{r}_{9}\frac{\partial{n}_F(\epsilon)}{\partial\epsilon}
e^{-i\Phi}\tilde{v}_{y}(\mathbf{r}_{9},\mathbf{r}_1)\tilde{v}_{x}(\mathbf{r}_{7},\mathbf{r}_8)
\tilde{g}^A(\mathbf{r}_8-\mathbf{r}_4;\epsilon)\tilde{g}^A(\mathbf{r}_4-\mathbf{r}_{9};\epsilon)\\\nonumber
&\times\tilde{g}^{R}(\mathbf{r}_1-\mathbf{r}_2;\epsilon)\tilde{g}^{A}(\mathbf{r}_4-\mathbf{r}_2;\omega-\epsilon)
\tilde{g}^{A}(\mathbf{r}_6-\mathbf{r}_4;\omega-\epsilon) \tilde{g}^R(\mathbf{r}_6-\mathbf{r}_7;\epsilon)
\sum_{N}\varphi_{N,0}(\mathbf{r}_2-\mathbf{r}_6)(\tilde{C}_{N}^R(\epsilon,\omega-\epsilon))^2\\\nonumber&\times
\left[\tilde{L}_{N}^{R}(\omega)(n_P(\omega)+n_F(\omega-\epsilon))+\tilde{L}_{N}^{A}(\omega)n_P(\omega)\right] +c.c.
\end{align}
Here $\ell_H=\sqrt{c/2eH}$ is the magnetic length for the $2e$ excitations in the Cooper channel, $\tilde{v}_x(\mathbf{r}_9,\mathbf{r}_1)=
\lim_{\mathbf{r}_{9}\rightarrow\mathbf{r}_1}\left(\boldsymbol{\nabla}_1^x/{2m}+ieH(y_1-y_2)/{4mc}-
\boldsymbol{\nabla}_{9}^x/{2m}-ieH(y_4-y_9)/{4mc}\right)$ is the velocity  written in its gauge invariant form, and $n_P(\omega)$ is the Bose distribution function. The phase $\Phi$  is the flux enclosed by the paths of all charged excitations:
\begin{align}\label{eq:3-flux}
\Phi=e\mathbf{H}\left[(\mathbf{r_4-r_1})\times(\mathbf{r_1-r_2})+(\mathbf{r_6-r_7})\times(\mathbf{r_7-r_4})+2(\mathbf{r_6-r_4})\times(\mathbf{r_4-r_2})\right]/2c.
\end{align}
All propagators of the collective modes ($\tilde{L}$ as well as $\tilde{C}$)  have the same  Landau level index. As we   show later, this is not always the case. Since all terms in the above equation are translational invariant (functions of the relative coordinates alone), we can rewrite the integral in terms of the relative momenta. Then, the spatial coordinates appearing in  the flux $\Phi$ and diamagnetic term become derivatives with respect to the  momenta:
\begin{align}\label{eq:3-Hall2C2}
&{j}_{NEW}^{y}=-i\frac{e^2E_x}{4\pi\nu\tau\sqrt{2\pi\ell_H^2}}\sum_{N}\int\frac{d\epsilon{d}\omega}{(2\pi)^2}\frac{d\mathbf{k}_1...d\mathbf{k}_{6}d\mathbf{q}}{(2\pi)^{3d}}\delta(\mathbf{k}_2-\mathbf{k}_3)\delta(\mathbf{k}_1-\mathbf{k}_6)\delta(\mathbf{k}_3+\mathbf{k}_4-\mathbf{q})
\delta(\mathbf{k}_5+\mathbf{k}_6-\mathbf{q})
\frac{\partial{n}_F(\epsilon)}{\partial\epsilon}\\\nonumber&\times
\exp\left\{-i\frac{eH}{2c}\left[\frac{\partial}{\partial\mathbf{k}_2}\times\frac{\partial}{\partial\mathbf{k}_3}+
\frac{\partial}{\partial\mathbf{k}_6}\times\frac{\partial}{\partial\mathbf{k}_1}+
2\frac{\partial}{\partial\mathbf{k}_5}\times\frac{\partial}{\partial\mathbf{k}_4}
\right]\right\}
\left(\frac{ik_3^{y}}{2m}-\frac{eH}{4mc}\frac{\partial}{\partial{k}_3^x}+
\frac{ik_2^{y}}{2m}+\frac{eH}{4mc}\frac{\partial}{\partial{k}_2^x}
\right)\\\nonumber&\times
\left(\frac{ik_1^{x}}{2m}-\frac{eH}{4mc}\frac{\partial}{\partial{k}_1^y}+
\frac{ik_6^{x}}{2m}+\frac{eH}{4mc}\frac{\partial}{\partial{k}_6^y}
\right)
\tilde{g}^A(\mathbf{k}_1;\epsilon)\tilde{g}^A(\mathbf{k}_2;\epsilon)\tilde{g}^{R}(\mathbf{k}_3;\epsilon)\tilde{g}^{A}(\mathbf{k}_4;\omega-\epsilon)\\\nonumber
&\times
\tilde{g}^{A}(\mathbf{k}_5;\omega-\epsilon) \tilde{g}^R(\mathbf{k}_6;\epsilon)
\sum_{N}\varphi_{N,0}(\mathbf{q})(\tilde{C}_{N}^R(\epsilon,\omega-\epsilon))^2
\left[\tilde{L}_{N}^{R}(\omega)(n_P(\omega)+n_F(\omega-\epsilon))+\tilde{L}_{N}^{A}(\omega)n_P(\omega)\right] +c.c.
\end{align}

The magnetic field enters $\hat{L}$ and $\hat{C}$ as $\Omega_c/T$ which is not necessarily small and, hence, we cannot expand in this parameter. In contrast, the flux can be expanded in powers of the magnetic field. Since each power introduces an additional derivative with respect to the quasiparticle momentum which can act either on the velocity vertex or the Green's functions, the small parameter emerging from the expansion is $\omega_c\tau\ll1$. Similar smallness is associated with the diamagnetic term. Nevertheless, the magnetic field entering via $\Phi$  cannot be neglected, as the zero order term vanishes. Actually,  extracting the magnetic field from the flux is the reason why the new contribution is of  the same order as the contribution corresponding to the diagram in Fig.~\ref{fig:diagrams}. In contrast, the contribution from Fig.~\ref{fig:new2} to the longitudinal conductivity (obtained by replacing $v_{y}$ by $v_{x}$ in the vertex) is smaller by a factor of $T\tau$ than all other terms described in Fig.~\ref{fig:diagrams}.   Following Ref.~\onlinecite{Khodas2003}, we can obtain all non-zero contributions arising from expansion of the flux to the first order in $H$. Then, we can integrate over the quasiparticle momenta, $\mathbf{k}_i$ and frequency, $\epsilon$.  Under the approximation of constant  density of states and velocity in the vicinity of the Fermi energy, we  get that the integral vanishes. Keeping corrections to this approximation, $\nu(\xi)\approx\nu+\nu'\xi\varepsilon_F$ and $v(\xi)=v_F+\xi/\varepsilon_F$,  results in a non-vanishing contribution to the Hall conductivity. Despite the smallness usually associated with these corrections, here it gives contribution to $\delta\sigma_{xy}$ comparable to all others:
\begin{align}\label{eq:3-Hall2C3}\nonumber
{j}_{NEW}^{y}&=-i\frac{e^3E_xH}{4\pi\ell_H^2c}\nu{D\tau^3}\sum_{N\geq0}\int\frac{d\epsilon{d}\omega}{2\pi}\frac{v_F^2}{d}
\left(\frac{1}{\varepsilon_F}+\frac{\nu'}{\nu_0}\right)\\
&\left\{(\tilde{C}_{N}^R(\epsilon,\omega-\epsilon))^2\left[(n_F(\epsilon-\omega)-n_F(\epsilon))\frac{\partial{n_P(\omega)}}{\partial\omega}\tilde{L}_{N}^{R}(\omega)-n_P(\omega)\frac{\partial{n}_F(\epsilon)}{\partial\epsilon}\tilde{L}_{N}^{A}(\omega)\right]\right\}-c.c
\end{align}
Further integration over the Bosonic frequency $\omega$ and summation over the Landau level $N$ is standard, and analytical solutions can be obtained in several limiting cases. The other part of the new contribution presented in Fig.~\ref{fig:new1}, gives exactly the same result. In the same way, we can derive the  contributions from the two Cooperons diagrams shown in Figs.~\ref{fig:diagrams}b,~\ref{fig:diagrams}e,~\ref{fig:diagrams}f,~\ref{fig:diagrams}i and~\ref{fig:diagrams}.j. While the expression corresponding to Figs.~\ref{fig:diagrams}b is identically zero, the rest of the terms are nonzero and their contributions are proportional to $\omega_c\tau$. However, the sum of these four diagrams vanishes.

As a representative example of the terms in the second group, we present the derivation of the Aslamazov-Larkin correction (see Fig.~\ref{fig:AL}). To keep our demonstration as simple as possible, we consider only part of the term (only contributions in which one propagator $L$ is retarded and the other is advanced):
\begin{align}\label{eq:3-ALFlux2}\nonumber
j_{AL}^{y}(\mathbf{r}_{\scriptscriptstyle1})&=-\frac{e^2E_x}{4\pi\ell_H^2}\int\frac{d\epsilon{d}\epsilon'd\omega}{(2\pi)^3}\sum_{N,M}\int{d\mathbf{r}_{\scriptscriptstyle2}...d\mathbf{r}_{\scriptscriptstyle12}}
e^{-i\Phi}\tilde{v}_{y}(\mathbf{r}_{12},\mathbf{r}_1)\tilde{v}_{x}(\mathbf{r}_{6},\mathbf{r}_7)
\tilde{g}^R(\mathbf{r}_{\scriptscriptstyle1}-\mathbf{r}_{\scriptscriptstyle2},\epsilon)
\tilde{g}^A(\mathbf{r}_{\scriptscriptstyle11}-\mathbf{r}_{\scriptscriptstyle2},\omega-\epsilon)\\\nonumber
&\times\tilde{g}^R(\mathbf{r}_{\scriptscriptstyle11}-\mathbf{r}_{\scriptscriptstyle12},\epsilon)
\tilde{g}^R(\mathbf{r}_{\scriptscriptstyle5}-\mathbf{r}_{\scriptscriptstyle6},\epsilon')\tilde{g}^A(\mathbf{r}_{\scriptscriptstyle5}-\mathbf{r}_{\scriptscriptstyle8},\epsilon')
\tilde{g}^R(\mathbf{r}_{\scriptscriptstyle7}-\mathbf{r}_{\scriptscriptstyle8},\epsilon')
\varphi_{N,0}(\mathbf{r}_{\scriptscriptstyle2}-\mathbf{r}_{\scriptscriptstyle5})\varphi_{M,0}(\mathbf{r}_{\scriptscriptstyle8}-\mathbf{r}_{\scriptscriptstyle11})\\
&\times \tilde{C}_N^{R}(\epsilon,\omega-\epsilon)\tilde{C}_M^R(\epsilon,\omega-\epsilon)
\tilde{L}_{N}^{R}(\omega)\tilde{L}_{M}^{A}(\omega)
\tilde{C}_N^R(\epsilon',\omega-\epsilon')\tilde{C}_M^R(\epsilon',\omega-\epsilon')F(\epsilon,\epsilon',\omega).
\end{align}
Here, $F(\epsilon,\epsilon',\omega)=\left[\tanh\left({\epsilon}/{2T}\right)-\tanh\left(({\epsilon-\omega})/{2T}\right)\right]\tanh\left(({\omega-\epsilon'})/{2T}\right) {\partial{n_P(\omega)}}/{\partial\omega}$, and the gauge invariant velocity $\tilde{v}$ was already defined in the previous example. The phase $\Phi$ is:
\begin{align}\label{eq:3-ALFlux22}
\Phi=\frac{e\mathbf{H}}{2c}\left[(\mathbf{r}_{\scriptscriptstyle11}-\mathbf{r}_{\scriptscriptstyle1})\times(\mathbf{r}_{\scriptscriptstyle1}-\mathbf{r}_{\scriptscriptstyle2})+(\mathbf{r}_{\scriptscriptstyle5}-\mathbf{r}_{\scriptscriptstyle6})\times(\mathbf{r}_{\scriptscriptstyle6}-\mathbf{r}_{\scriptscriptstyle8})+2(\mathbf{r}_{\scriptscriptstyle2}-\mathbf{r}_{\scriptscriptstyle5})\times(\mathbf{r}_{\scriptscriptstyle5}-\mathbf{r}_{\scriptscriptstyle8})+2(\mathbf{r}_{\scriptscriptstyle8}-\mathbf{r}_{\scriptscriptstyle11})\times(\mathbf{r}_{\scriptscriptstyle11}-\mathbf{r}_{\scriptscriptstyle2})\right].
\end{align}
The first two terms in Eq.~\ref{eq:3-ALFlux22} correspond to the magnetic fluxes accumulated in the triangles $(\mathbf{r}_{\scriptscriptstyle1},\mathbf{r}_{\scriptscriptstyle2},\mathbf{r}_{\scriptscriptstyle11})$ and $(\mathbf{r}_{\scriptscriptstyle5},\mathbf{r}_{\scriptscriptstyle6},\mathbf{r}_{\scriptscriptstyle8})$, respectively. One may check that the contributions to the transverse current obtained by expanding the fluxes from these two triangles or the diamagnetic terms vanish. Therefore, the integration over the coordinates of the two triangles can be done with the quasiparticle Green's functions taken at $\mathbf{H}=0$. After integrating over the quasiparticles degrees of freedom, the triangles $(\mathbf{r}_{\scriptscriptstyle1},\mathbf{r}_{\scriptscriptstyle2},\mathbf{r}_{\scriptscriptstyle11})$ and $(\mathbf{r}_{\scriptscriptstyle5},\mathbf{r}_{\scriptscriptstyle6},\mathbf{r}_{\scriptscriptstyle8})$ become proportional to gradients acting on the propagators in the particle-particle channel. Using the remaining two fluxes, corresponding to the triangles $(\mathbf{r}_{\scriptscriptstyle2},\mathbf{r}_{\scriptscriptstyle5},\mathbf{r}_{\scriptscriptstyle8})$ and $(\mathbf{r}_{\scriptscriptstyle2},\mathbf{r}_{\scriptscriptstyle8},\mathbf{r}_{\scriptscriptstyle11})$ , the expression for the current can be written in the following way:
\begin{align}\label{eq:3-ALFlux3}
j_{AL}^{y}&=-\frac{e^2E_x}{8\pi^2\ell_H^2}\nu^2\tau^4\int{d\epsilon{d}\epsilon'd\omega}\int{d\mathbf{r}}\sum_{N,M}
\left[2D\left(\frac{\partial}{\partial{y}}-\frac{ie{H}x}{c}\right)\varphi_{N,0}(\mathbf{r})\right]
\left[2D\left(\frac{\partial}{\partial{x}}+\frac{ieHy}{c}\right)\varphi_{M,0}(\mathbf{r})\right]\\\nonumber
&\times
\tilde{C}_N^R(\epsilon,\omega-\epsilon)\tilde{C}_M^R(\epsilon,\omega-\epsilon)
\tilde{L}_{N}^{R}(\omega)\tilde{L}_{M}^{A}(\omega)
\tilde{C}_N^R(\epsilon',\omega-\epsilon')\tilde{C}_M^R(\epsilon',\omega-\epsilon')F(\epsilon,\epsilon',\omega).
\end{align}
Let us define the velocity operator, $V_i=2D\left({\nabla_i}-{ie(\mathbf{H}\times\mathbf{r})_i}/{c}\right)$, of an auxiliary particle with a mass equal to $1/2D$. The integral over the coordinate corresponds to the matrix element of the velocity operators $\langle{N,0}|V_iV_j|M,0\rangle$, where $|M,0\rangle=\varphi_{M,0}$ is the quantum state of the particle in the $M$ Landau level and zero angular momentum in the $z$-direction. Using the known properties of the Laguerre polynomials, the matrix element can be written as $\langle{N,0}|V_iV_j|M,0\rangle=4ieD^2H[(N+1)\delta_{N,M-1}+(-1)^{i+j}(M+1)\delta_{M,N-1}]/c$. Finally, the contribution to the current acquires the form:
\begin{align}\label{eq:3-ALFlux4}
&j_{AL}^{y}=i\frac{e^3E_xH}{2\pi^2\ell_H^2c}\nu^2D^2\tau^4\int{d\epsilon{d}\epsilon'd\omega}\sum_{N=0}^{\infty}
(N+1)\tilde{C}_N^R(\epsilon,\omega-\epsilon)\tilde{C}_{N+1}^R(\epsilon,\omega-\epsilon)
\\\nonumber&\times{\tilde{C}}_N^R(\epsilon',\omega-\epsilon'){\tilde{C}}_{N+1}^R(\epsilon',\omega-\epsilon')
\left[\tilde{L}_{N}^{R}(\omega)\tilde{L}_{N+1}^{A}(\omega)-\tilde{L}_{N+1}^{R}(\omega)\tilde{L}_{N}^{A}(\omega)\right]
F(\epsilon,\epsilon',\omega).
\end{align}
\end{widetext}
In the derivation of the new contribution discussed previously, we had to
keep corrections to the constant density of states but could set the other
small parameter $\varsigma=0$. Here we must keep $\varsigma$ non-zero, while
assuming $\nu(\epsilon)$ to be constant. The vanishing of $\delta\sigma_{xy}$
when both $\nu(\epsilon)=const$ and $\varsigma=0$ occurs because the Hall
conductivity is zero in a particle-hole symmetric system. Consequently, we
found that the Aslamazov-Larkin contribution to $\delta\sigma_{xy}$ is
proportional to $\Omega_c\varsigma$.

In the same way, we can derive the remaining parts of the Aslamazov-Larkin
terms, density of state corrections to the conductivity  as well as the three Cooperons diagrams presented in Figs.~\ref{fig:diagrams}c and~\ref{fig:diagrams}d. The  contributions of the first two to the Hall conductivity  are given in Eqs.~\ref{eq:3-j_AL} and~\ref{eq:3-Hall3C}. Examining Fig.~\ref{fig:diagrams}c, one can see that it is a mirror  image of Fig.~\ref{fig:diagrams}d. Therefore, they acquire opposite signs as a result of turning the current using the magnetic field, and their sum is identical zero.

\end{document}